# An Attempt to Catch Up with JIT Compilers

## The False Lead of Optimizing Inline Caches

Aurore Poirier[a], Erven Rohou[a], and Manuel Serrano[b]

a University of Rennes - Inria - CNRS - IRISA, France

b Inria - University of Côte d'Azur, France

**Abstract**

**Context** Just-in-Time (JIT) compilers are able to specialize the code they generate according to a continuous profiling of the running programs. This gives them an advantage when compared to Ahead-of-Time (AoT) compilers that must choose the code to generate once for all.

**Inquiry** Is it possible to improve the performance of AoT compilers by adding Dynamic Binary Modification (DBM) to the executions?

**Approach** We added to the Hopc AoT JavaScript compiler a new optimization based on DBM to the inline cache (IC), a classical optimization dynamic languages use to implement object property accesses efficiently.

**Knowledge** Reducing the number of memory accesses as the new optimization does, does not shorten execution times on contemporary architectures.

**Grounding** The DBM optimization we have implemented is fully operational on `x86_64` architectures. We have conducted several experiments to evaluate its impact on performance and to study the reasons of the lack of acceleration.

**Importance** The (negative) result we present in this paper sheds new light on the best strategy to be used to implement dynamic languages. It tells that the old days where removing instructions or removing memory reads always yielded to speed up is over. Nowadays, implementing sophisticated compiler optimizations is only worth the effort if the processor is not able by itself to accelerate the code. This result applies to AoT compilers as well as JIT compilers.

## 1 Introduction

The fastest contemporary JavaScript implementations use JIT compilers [27]. Their ability to continuously tune the code they generate using dynamic profiling information enables performance that seemed impossible with the first JavaScript implementations. However, JIT compilers may not be desirable or simply not available in some contexts, for instance if programs are to be executed on platforms with too limited resources or if the architecture forbids dynamic code generation. Ahead of time (AoT) compilers offer a response to these situations.

`Hopc` [25] is an AoT JavaScript-to-C compiler. Its performance is often in the same range as that of the fastest JIT compilers but its impossibility to adapt the code executed at runtime seems a handicap for some patterns and benchmarks [27]. This paper presents an experiment we conducted to combine AoT and JIT compilation in order to take benefit of both approaches. More precisely, we added a new *dynamic binary modification* (DBM) optimization [31] inherited from the early implementations of dynamic languages [10] to the `hopc` runtime system. It optimizes one step further, at runtime, the seminal code of the inline caches [8] generated by the static compiler. It removes up to two memory accesses for each successful inline cache hit by modifying the code at runtime.

In this paper we show how we implemented the DBM optimization in the context of the `hopc` compiler and we show that, despite effectively removing memory accesses (as expected), it does not accelerate executions.

Some works have shown that many preconceptions about performance [30] have to be reevaluated in light of the continuous evolution of hardware architectures [24]. For instance, *what is the cost of a conditional expression if the micro-architecture supports efficient branch prediction?* Or also, *how important is it to rely on a precise static register allocation if the micro-architecture supports many more registers than the instruction set exposes?* More generally, speculation, caches, pre-fetchting, and all the other techniques micro-architectures deploy to accelerate executions, make it difficult for compiler optimizers to have a clear model of the most efficient code they should generate. This is what happens with the optimization presented in this paper: it produces better code, it reduces the number of memory accesses, but it optimizes a pattern that micro-architectures are apparently already able to optimize by themselves! This is the reason why the optimization presented in this paper does not accelerate executions. Although, this is a deceptive result for the DBM, we believe that it is valuable to publish, at least to prevent other researchers from entering the same dead-end, and also, of course, because the techniques we developed to combine AoT and JIT compilation could be useful and beneficial in other contexts.

The rest of the paper is organized as follows. Section 2 introduces the inline caching technique used by all optimizing JavaScript implementations. This section shows the differences between the binaries JIT and AoT compilers can generate. Section 3 presents the dynamic optimization we added to `hopc` in order to make its inline caches similar to that of JIT compilers. Sections 4 present the performance evaluation and explain the reasons for the lack of speedup.

Section 5 presents JavaScript implementation and DBM performance related works. Section 6 concludes.

## 2 Inline Caching

JavaScript is a prototype-based object-oriented programming language. In this programming model, objects are associative tables and fetching a property involves a lookup in the object itself and, if the property is not directly found, in its *prototype* chain. If implemented naively, these lookup operations are slow and constitute a performance bottleneck. Fortunately, techniques invented for Smalltalk and Self in the 80's, named *inline caches* and *hidden classes* [8, 10], apply to JavaScript as well. They are a key component of all fast modern JavaScript implementations.

The *inline cache* (IC) technique relies on the observation that considering a property read "`obj.prop`" in a program, the successive values of `obj` on that same point are likely to share the same *shape*, and so the property `prop` is also likely to be stored at the same offset in these `obj` values.

To take benefit of this observation, the internal implementation of objects is augmented with a so-called *hidden class*. When a new property is added to or removed from an object, its hidden class is updated. A hidden class stores the set of properties an object holds and where these properties are stored in the object itself.

Object accesses are optimized by remembering, in an IC, the hidden class of the previously accessed object and by comparing it to the one of the object being accessed. On a match, that is when the two classes are the same, the location of `prop` in `obj` is known and can be directly read without any lookup. In C this optimization can be implemented as shown in Listing 1.

**Listing 1** Code of an inline cache access.

```
1 if (obj->hclass == icache.hclass) {        // cache comparison
2   prop = *(obj + icache.offset);           // cache hit
3 } else {
4   prop = cache_miss(obj, "prop", &icache); // miss
5 }
```

At line 1, `obj`'s hidden class is compared to that of the previous object used at that program point (note that there is one IC per static property read of the program). If the hidden classes are the same (line 2), the memory representations of the objects are the same too, hence the cached offset is used and no lookup is necessary. Otherwise, since the two objects differ (line 4), a slow lookup is needed. This lookup implements the JavaScript semantics [11]. It searches for `prop` in `obj`. If `prop` is found in `obj`, `icache` is updated. For the sake of simplicity, we do not show the implementation of `cache_miss`.

This optimization is effective if most of the objects share the same representation and consequently the same hidden class, and if the property is found in the object itself instead of in the prototype chain. Refined inline caching techniques have been proposed [14], to handle other situations. In this study, we only consider simple IC.

All fast modern JIT-based JavaScript implementations use inline caches (V8, JavaScriptCore, SpiderMonkey). Pizlo's blog [23] details the implementation of JavaScriptCore (JSC). It shows the actual assembly code generated for a property access (see Listing 2). Lines 1, 2 implement the hidden class comparison (similar to line 1 of Listing 1). Line 3 loads the property on a cache hit. On a cache miss, immediate values in the instructions at lines 1 and 3 (`0x125` and `0x18`) have to be updated to the new hidden class and property offset. JavaScriptCore also generates a code with a multi-byte `nop` instruction to reserve space for later dynamic modifications [23].

**Listing 2** JavaScriptCore generated assembly code.

```
1 cmp $0x125, (%rax)       # compare the hidden class
2 jnz .slow_access         # cache misses
3 mov 0x18(%rax), %rax     # load the property
4 nop 0x200(%rax)          # free space
```

Listing 3 shows the code `gcc` generates for our C sequence. Line 1 reads the hidden class contained in the IC. At line 2 the object's hidden class is loaded. In the assembly excerpt of JavaScriptCore, it was assumed to be already loaded in `%rax` and this code to do so is not shown. Lines 3 and 4 map one-to-one to lines 1 and 2 of the JSC code (Listing 2). They implement the hidden class comparison. Line 5 gets the property offset from the inline cache, and line 6 finally loads the data into `rax`.

**Listing 3** `gcc` generated assembly code edited for the sake of the comparison with JavaScriptCore.

```
1 mov 0x1000(%rip), %rbx  # load icache.hclass
2 mov 0x30(%rbp), %rax    # load object->hclass
3 cmp %rbx, %rax          # compare the hidden classes
4 jne .slow_access        # cache misses
5 mov 0x101c(%rip), %rax  # load the offset
6 mov (%rdi,%rax,8), %rax # load the property
```

Lines 3, 5 and 6 of the C generated code are where the difference with the JSC code might impact the performance. The difference comes from C using global variables to store cached property offsets (the variable `icache` in the excerpt) instead of, as JSC, directly encoding the property offset in the new assembly instructions it generates on cache misses. Consequently, JSC uses one assembly instruction that executes a single memory access, while the `gcc` code uses two instructions and more importantly, two memory loads. Object accesses are ubiquitous and one of the most frequent operations. As such, their performance is crucial for the global performance. The `hopc` compiler, which generates C code presented in the current section, suffers from a disadvantage when compared to JIT compilers. We have developed a technique that relies on dynamic binary modification (DBM) in the context of C compiled code to let it eventually generate the same assembly code as JIT compilers for accessing properties.

☞ As we show in Section 4, this new DBM optimization *succeeds* at eliminating the extra memory reads of `hopc`, but as we will also see in that section, counterintuitively, removing them *does not* accelerate executions.

## 3 Dynamic Binary Modification and C Generated Code

In this section, we present the techniques we developed to remove the extra memory read imposed by the use of a C global variable in the IC sequences (lines 5 and 6, Listing 3). These techniques rely on dynamic modification of the assembly instructions generated by optimizing C compilers.

Modifying the C generated IC sequences at runtime raises two main challenges.

1. How to discover the addresses of the instructions corresponding to the IC hit in the assembly code?
2. How to decode the assembly instructions to prepare their dynamic binary modification?

There is no portable way to associate an assembly instruction with its origin in a C source file. Even worse, because C compilers use complex and sophisticated optimizations, there is no guarantee that such mappings always exist. So, we developed a technique that *heuristically* enables the dynamic finding and modification of assembly instructions. This technique is conservative and sound. It is not guaranteed to find all the possible IC sequences, but it never modifies instructions not involved in a property access. The experimental report presented in Section 4 shows that for "gcc -O3" generated code, the technique we developed effectively discovers 96.9 % of all the IC sequences.

### 3.1 Modifications to the Generated C Code

We modified hopc's C code generator to associate a unique C label with each IC. These labels are used at runtime to retrieve the addresses of the assembly instructions to be modified on cache misses. The "399" in the &&IC_LBL399 label of Listing 4 is a unique identifier denoting this very IC. Listing 5 shows the code generated by gcc when IC labels are added.

**Listing 4** C label added to the IC sequence to enable the DBM of the "cache hit" expression.

```
if (obj->hclass == icache.hclass) {        // cache comparison
IC_LBL399:                                 // per inline cache label
  prop = *(obj + icache.offset);           // cache hit
} else {
  prop = cache_miss(obj, "prop", &icache,  // cache miss
                    &&IC_LBL399);          // the label address
}
```

**Listing 5** Possible C generated assembly code after adding the IC_LBL399 label.

```
  mov 0x1000(%rip), %rbx  # load icache.hclass
  mov 0x30(%rbp), %rax    # load object->hclass
  cmp %rbx, %rax          # compare the hidden classes
  jne .slow_access        # cache misses
IC_LBL399:
  mov 0x101c(%rip), %rax  # load the offset
  mov (%rdi,%rax,8), %rax # load the property
```

To retrieve the address of the C label at runtime, `hopc` uses the `gcc` extension *Labels as Values*[1] also supported by `clang`, by the Intel compilers `icc`, and by oneAPI `icx`. This C extension provides the `&&` operator that gives the address of a label (Listing 4, line 6).

On a cache miss, the `cache_miss` function first searches for the property in the object, and then in its prototype chain if not found. If it finds the property in the object, before returning the property's value it scans the assembly instructions from address `&&IC_LBL399` for a possible code modification opportunity. If it recognizes the instructions, it stores relevant information for future possible misses so that only the first inline cache miss incurs a consequent overhead. The subsequent misses are faster to handle, limited to writing proper values to already computed addresses. Failure to recognize the IC sequence is also stored in a cache, in order to avoid analyzing again the same assembly instructions on next misses.

We implemented this technique on `x86_64` architectures. These architectures support variable-size instructions and many addressing modes, which makes decoding instructions tedious. To avoid implementing a full `x86_64` decoder, we used the Capstone [7] tool, which is lightweight multi-architecture disassembly framework. On a first cache miss, the `cache_miss` function uses Capstone to disassemble a small window of instructions at the address of the IC associated label.

### 3.2 Assembly Instruction Research

The next step is to find the assembly instructions that implement the property read of the IC code sequence (lines 6 and 7, Listing 5). As IC assembly sequences are generated by optimizing C compilers which combine many optimizations, it happens that sequences greatly differ from one context to another. Consequently, it also happens that the `hopc` dynamic search for assembly instructions fails. In such a case, no dynamic optimization is applied to the corresponding IC and the program executes as if only statically optimized.

The dynamic code analysis scans the instructions from the address of the label associated to the IC (`IC_LBL399` in the example of Listing 5). It stops when it finds the read instructions (lines 6 and 7 in the same example) or when it hits an instruction it does not recognize, for instance a branching instruction such as a `jmp` or a `call`. The main reasons for this process to fail finding the read instructions are the following:

- the C compiler generated an unusual instruction sequence that the analyzer does not recognize;
- the instructions have been moved far away from the IC label;
- the memory access is not RIP-Relative (see end of Section 3.2), which is, according to our observations, the most frequent cause of failures.

To consider an assembly instruction as a candidate for dynamic modification, it should:

[1] https://gcc.gnu.org/onlinedocs/gcc/Labels-as-Values.html, accessed on 2025-01-31.

1. be a `mov`;
2. read from memory;
3. store the result to a register;
4. read from the IC property offset;
5. read the offset of an IC sequence.

The first four conditions are mere syntactic checks. To implement the fifth one, the DBM optimizer receives as an extra parameter a pointer to the IC structure that it compares to the memory location accessed by the assembly instruction (`0x101c(%rip)` in line 6 of Listing 5). If the two pointers differ, the optimizer bails.

The analysis of IC hit sequences takes place during the execution, on IC misses. This analysis is ignorant of the register values and the content of the memory, except for the *Instruction Pointer* (`%rip` in `x86_64`). Thus, only RIP-Relative memory accesses are rewritten. Thankfully, on `x86_64`, popular optimizing C compilers generate RIP-Relative instructions for the sake of code size and code locality. Implementing the optimization described in this paper for architectures or compilers that do not rely on some sort of program counter relative addressing would be more difficult.

Once the first instruction of the property read sequence is found (Listing 5, line 6) it is modified so that the memory load instruction is replaced with an immediate load whose value corresponds to the offset of the property. Listing 6 shows the modified code. The instruction at line 6 has been replaced with a `mov` and the one at line 7 has been replaced with a `nop`. We refer to this transformation as the `-O1` optimization level.

**Listing 6** IC sequence of Listing 5 after dynamic `-O1` optimization.

```
1   mov 0x1000(%rip), %rbx  # load icache.hclass
2   mov 0x30(%rbp), %rax    # load object->hclass
3   cmp %rbx, %rax          # compare the hidden classes
4   jne .slow_access        # cache misses
5 IC_LBL399:
6   mov 0x3, %rax           # load the offset
7   nopl (%rax)             # NOP Padding
8   mov (%rdi,%rax,8), %rax # load the property
```

When the IC assembly instruction sequence has a shape similar to that of Listing 5 (*i.e.*, the second instruction in the sequence is a `mov` reading the register written by the first one), another optimization is possible: as the assembly instruction immediately following the offset load fetches the property from the object (as line 8 in Listing 6), the two instructions can be combined into a single instruction (see Listing 7). The constant immediate value is propagated from the first instruction into the second, and the first instruction is eventually removed. We refer to this transformation as the `-O2` optimization level.

The `-O2` optimization is more involved than `-O1`. To apply it, the second instruction must read the memory from a register containing the object location (`%rdi` in Listing 5). It must address the memory using the property cache offset, *i.e.*, the register written by the first instruction (`%rax` in Listing 5), possibly with an offset multiplier to match the word size (8 in Listing 5) and possibly a constant offset as defined by

x86_64 addressing modes [15, Vol. 2A, Section 2.1.3]. Finally, it must store the value in a register. If such an instruction is found, since the value of the first `mov` is known, it can be removed and replaced by pre-computing a fixed offset as in Listing 7 (here `0x18` $= 3 \times 8$ if we take the third field in the memory layout of the object's data).

**Listing 7** IC sequence of Listing 5 after dynamic `-O2` optimization.

```
  mov 0x1000(%rip), %rbx   # load icache.hclass
  mov 0x30(%rbp), %rax     # load object->hclass
  cmp %rbx, %rax           # compare the hidden classes
  jne .slow_access         # cache misses
IC_LBL399:
  mov 0x18(%rdi), %rax     # load the property
  nopl 0L(%rax)            # NOP Padding
```

For the sake of soundness, the `-O2` optimization also requires the register written to by the second instruction to be the same as the one written by the first (`%rax` in the example). This guarantees that the destination register of the first instruction is dead, hence the instruction is useless. Preserving soundness while relaxing this constraint would require a more complex register liveness analysis that we did not implement. When the registers used by the two instructions differ, the `hopc` DBM conservatively limits itself to the `-O1` optimization level.

### 3.3 Practical DBM Constraints

In practice, the implementation of DBM on C generated IC must address three last considerations:

1. system memory protections must be removed;
2. `x86_64` variable size instructions must be handled correctly;
3. some immediates do not fit in a `x86_64` instruction.

Generally, operating systems remove the write permission to memory pages that contain executable code. The DBM requires this permission to be granted. On Unix system, page protections can be overridden with the `mprotect` system call on a per page basis. As system calls are expensive (thousands of cycles and possibly even worse if we consider cache pollution footprint [28]), `hopc` stores un-protected pages in a dedicated cache to execute only one `mprotect` per page for a whole execution.

As `x86_64` uses variable length instructions, it happens that the DBM optimization replaces assembly instructions with shorter ones. In such a situation, it inserts a multi-byte `nop` assembly instruction padding. We observed that these `nop` instructions do not impact the overall execution times but they change the number of executed instructions. That is, if two instructions of an IC sequence are replaced with two shorter instructions *and* a padding `nop`, then one additional assembly instruction is executed. The hardware register counting the number of instructions will count this difference although these additional nop instructions do not impact the performance (`nop` instructions are detected in the early stages of the processor pipeline and quickly eliminated). This has to be considered when evaluating the impact of the optimization.

A last constraint is that no `x86_64` instruction can load an immediate bigger than 4 bytes [15, Vol. 2B, MOV]. It requires a check on the IC offset size. A consequence is that the IC hidden class read can not be rewritten, as the actual implementation use pointer (which are 8 byte long on a 64 bit architecture).

### 3.4 Summary by Example

Let us conclude this section with an example of an execution. Let us consider the example of Listing 8. It reads successively the `prop` property of the three objects. The first two share the same hidden class, the third one is different. The example executes the following steps:

1. an access to an object with a first hidden class $\mathscr{C}_0$ whose property is located at offset 24 (`0x18`, which corresponds to the $3^{rd}$ field of a structure on a 64-bit architecture);
2. a second access to another object with the same hidden class $\mathscr{C}_0$;
3. an access to an object with another hidden class $\mathscr{C}_1$ whose property `prop` is located at offset 32 (`0x20`).

**Listing 8** Example of polymorphic JavaScript memory access.

```
1 arr = [{a: 13, prop: 12}, {a: 13, prop: 37}, {b: "b", c: 4.2, prop: 2}];
2
3 for(const element of arr) {
4   content = element.prop;
5   ... content ...
6 }
```

When executing line 4 of Listing 8 for the first time, the IC missed because the cache is not populated yet. The test at line 1 of the C code of Listing 4 or lines 1-4 of the generated assembly code (repeated in Listing 9 for easier reading) fails. So, the program calls the `cache_miss` routine. It finds the `prop` property directly in the object, at offset `0x18`. As this sequence has not been already analyzed, it is first disassembled and the instructions starting at line 5 of Listing 9 are inspected. This instruction is successfully recognized by the DBM optimization:

- it is a `mov` instruction;
- it reads from memory (`0x101c(%rip)`);
- it stores the result in a register (`%rax`);
- it reads from the IC property offset, using an RIP-Relative read.

This is enough for `-O1` optimization level, but the analysis continues to line 6 of Listing 9 to discover that `-O2` can be applied too, as the next instruction meets all the required criteria:

- it reads from the previous instruction written register (the first `%rax` occurrence);
- it reads from the register containing an object location (`%rdi`);
- it writes to the same register as the first instruction (second occurrence of `%rax`).

**Listing 9** Assembly and binary of inline cache sequence.

```
1 48 8b 1d 00 10 00 00  mov    0x1000(%rip),%rbx
2 48 8b 45 30           mov    0x30(%rbp),%rax
3 48 39 d8              cmp    %rbx,%rax
4 0f 85 00 00 00 00     jne    .slow_access
5 48 8b 05 1c 10 00 00  mov    0x101c(%rip),%rax
6 48 8b 04 c7           mov    (%rdi,%rax,8),%rax
7 48 89 45 c8           mov    %rax,-0x38(%rbp)
```

Since the destination register is `%rax`, the object is located in `%rip` and the accessed field is at offset `0x18`, the instruction that needs to be written is

`mov 0x18(%rdi), %rax`.

It is a 7-byte long instruction (line 5 of Listing 10). The original instructions spanned over 11 bytes (lines 5 and 6 of Listing 9). There is enough space to write the instruction so the DBM modifies the program and uses a 4 byte long `nop` instruction for padding (see Listing 10).

**Listing 10** Assembly and binary of inline cache sequence after DBM.

```
1 48 8b 1d 00 10 00 00  mov    0x1000(%rip),%rbx
2 48 8b 45 30           mov    0x30(%rbp),%rax
3 48 39 d8              cmp    %rbx,%rax
4 0f 85 00 00 00 00     jne    .slow_access
5 48 8b 87 18 00 00 00  mov    0x18(%rdi),%rax
6 0f 1f 00 00           nopl   0L(%rax)
```

As this is the first use of this IC, the pages containing the IC sequence is unprotected so a call to `mprotect` is executed and the address of the page is stored in the memory protection cache. In addition, the very instruction containing the value of the offset of the IC is stored in memory so that when another hidden class is used on that IC sequence, only this instruction will be updated. This prevents from rewriting the whole IC sequence.

On the second execution of line 4 of Listing 8, the two hidden classes match so the test line 1 of Listing 4 succeeds and optimized IC code is used: the single instruction

`mov 0x18(%rdi), %rax`

is executed to read the property from the second object.

On the third execution of line 4 of Listing 8, a cache miss occurs. The `cache_miss` function uses the information it computed during the first miss to retrieve the information about the instructions to be modified to update the property offset. In our example, its writes `20 00 00 00` over the 4 last bytes of line 5 from Listing 10.

This concludes this section where we exposed the techniques we developed to apply DBM to C generated code and how we used it to optimize IC generated by the `hopc` AoT JavaScript compiler. We presented two levels of optimization. Their applicability depends on a local analysis of the assembly sequence generated by the C compiler for each inline cache. In the next section we evaluate the impact of the optimization.

# 4 Experimental Results

In this section we present the experiment we conducted to measure the impact of the DBM optimizations. We answer the following research questions:

*RQ*1 How effective are the techniques of Section 3 for detecting and modifying IC code sequences?

*RQ*2 What is the impact of the DBM optimizations on instruction count?

*RQ*3 How effective are the DBM optimizations to remove memory reads?

*RQ*4 What is the overall performance impact of the DBM optimizations?

## 4.1 Experimental Setup

For this experiment we used jsbench [26], a collection of widely used JavaScript benchmarks. They are made of a loop executing several times the same algorithm. In order to eliminates warm up effects, we tuned the loops so that each benchmark executes in no less than 10 s on our fastest platform (Golden Cove cores of Alder Lake architecture).

We compiled all the programs with the `hopc` option `--js-cspecs-get "(imap)"` which enables monophormic IC only. `Hopc` compiles JavaScript to C files that are then compiled with `-O3` optimizations. The tools used were `gcc` version 14.2.1 20240805, `hopc` commit `cae6066ed` for original version and commit `f9442a5fc` for DBM augmented version, and finally `perf` version 6.10-1 for measurements. Benchmarks were pinned on all CPU cores excluding the first physical core, which is in charge of managing IRQs, to avoid interferences. We executed the programs on two different machines for a total of three different architectures:

- a Dell Precision 3571 having an Intel® Core™ i7-12700H (2022) made of two different types of cores:
  - Golden Cove (performance);
  - Gracemont (efficiency);
- a custom desktop equipped with an AMD Ryzen 7 5700X made of:
  - Zen 3 cores.

Each benchmark has been executed 50 times. Address Space Layout Randomization (ASLR) and CPU frequency scaling were kept as their default values in order to get results reflecting normal usages.

The label `-O1` and `-O2` refer to the optimizations presented in Section 3. The label `-O0` refers to executions that do not modify instructions dynamically.

## 4.2 Validity of the Experimental Setup

It is well known that even the smallest modification of a program or its execution environment might impact noticeably its performance. For instance, changing the link ordering of modules, or declaring an extra environment variable may change the cache behavior and may randomly accelerate or slow down a program [18]. To

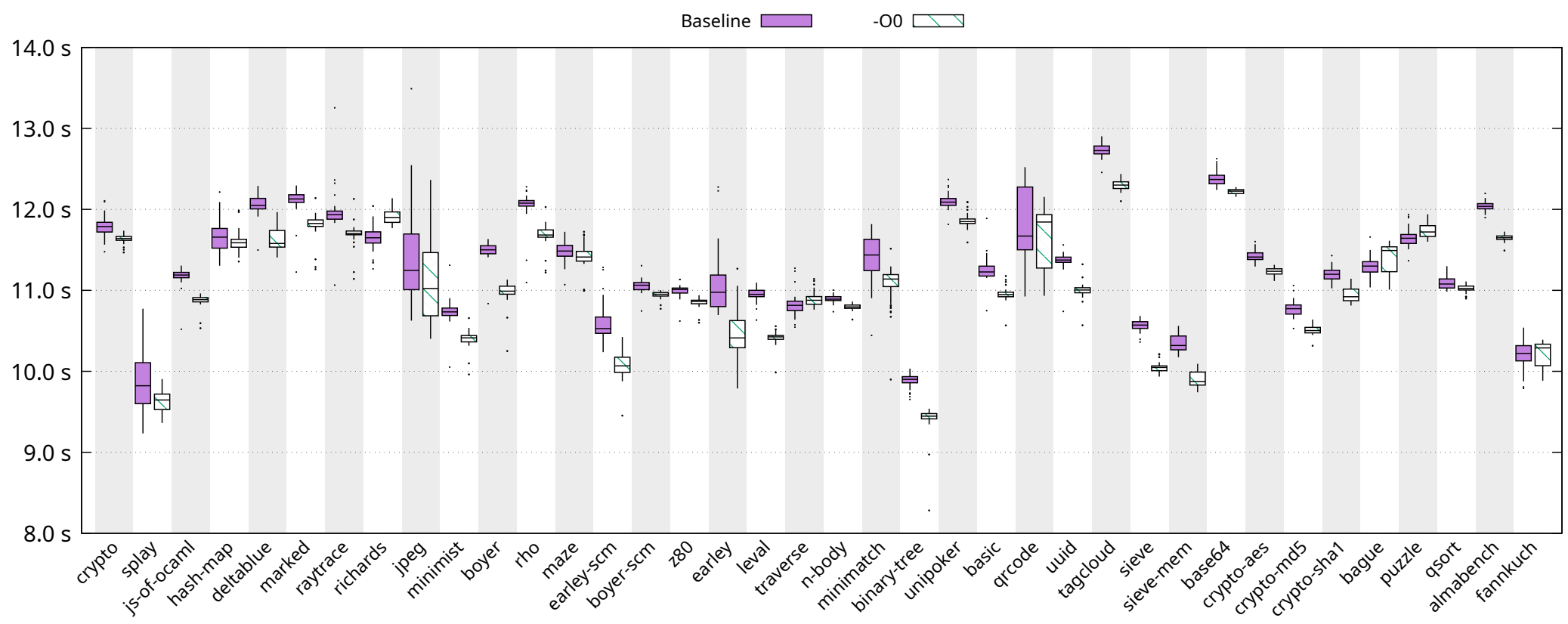

**(a)** On Golden Cove cores.

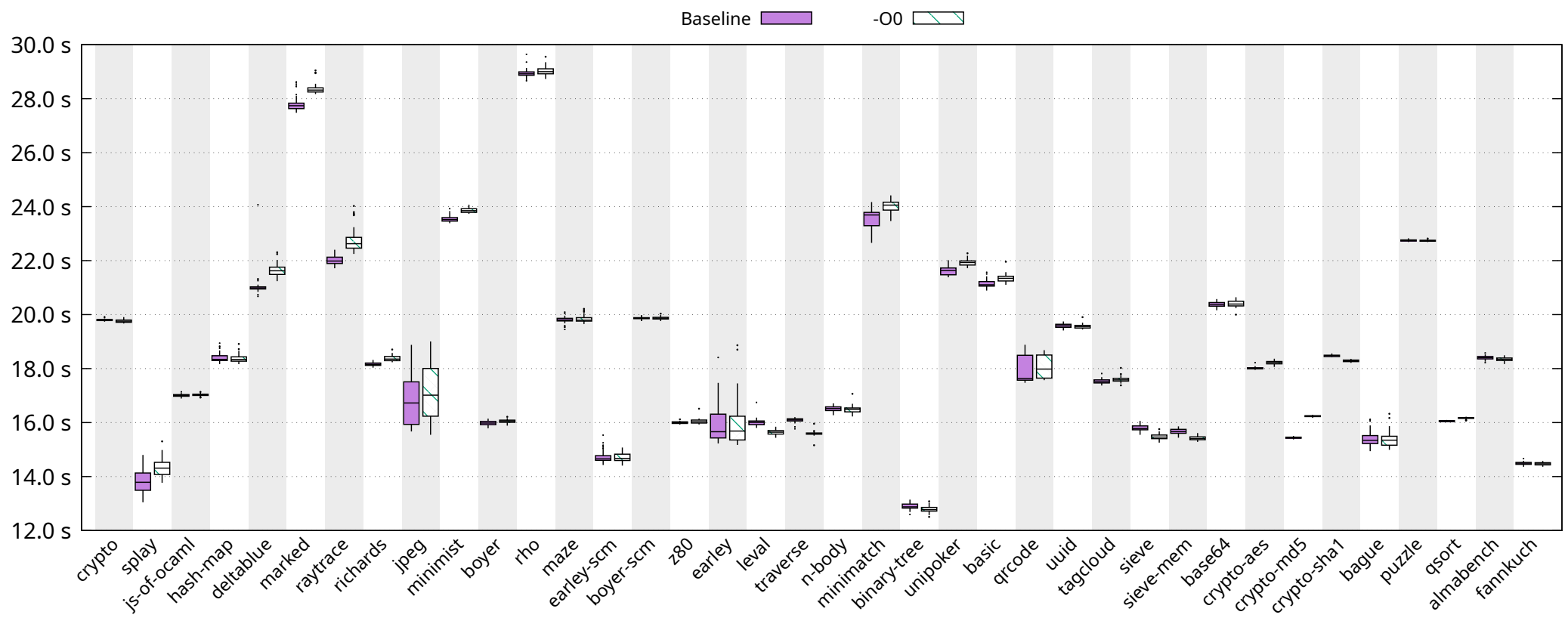

**(b)** on Gracemont cores.

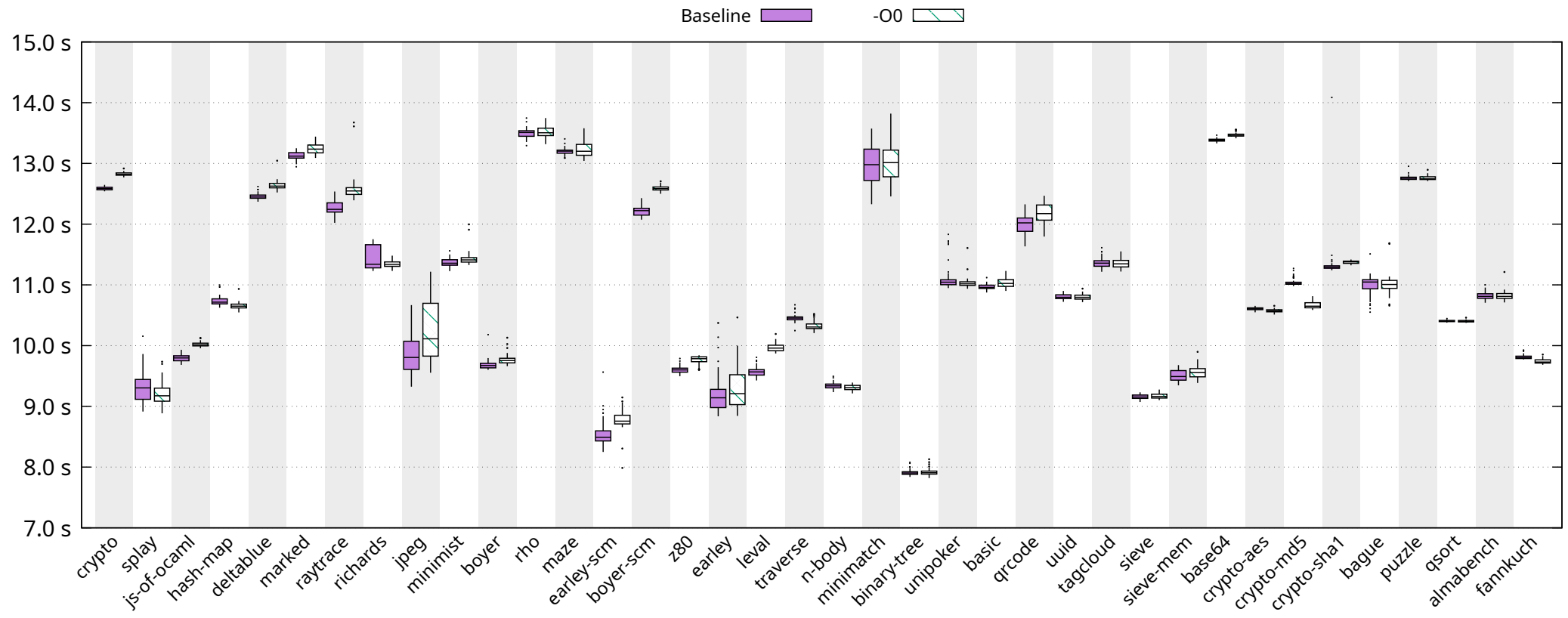

**(c)** on Zen 3 cores.

**Figure 1** Cost of adding the unexecuted DBM code in the benchmarks. No dynamic binary modification is ever executed for that experiment. Lower is better. Linear scale is used.

evaluate this effect, we measured the impact of adding the code needed for the DBM optimizations by comparing the execution times of `-O0` optimization and the execution times of the genuine `hopc` compiler that we label *baseline* in Figure 1. This experiment shows that adding *unexecuted* code (`-O0` versions) to programs impacts their performance by ±5 %. This conforms to the results presented in the literature about execution variability [9, 18].

Unlike a dynamic randomization approach [9] that may not suit DBM optimizations, we took a stabilization approach to get statistically sound results. Apart from the "baseline" version used in this section only, all versions described in Section 4 were executed using a single binary program. DBM optimizations are turned on and off by reading, at the initialization of the programs, the value of a shell environment variable. This environment variable always holds a value of exactly one character so that all stacks of all executions are aligned similarly. This ensures that performance differences we measure are not due to artifacts of the compilation or link processes.

Since not all factors from the system and CPU can be controlled, each measurement has been run 50 times. This high number of runs allows experiment to not fall in local extrema due to unaccounted for conditions. Keeping ASLR, allows bringing a form of stabilization as Curtsinger [9] did while still being able to apply DBM (addresses changing only when starting a new run).

Moreover, the experiments presented in this paper have been run on air cooled setups, providing a low thermal inertia, hence allowing to reach stable CPU temperatures and frequencies in fraction of seconds (over several hours or days of runs).

### 4.3 *RQ*1 How Effective Is the IC Sequence Detection?

Figure 2 shows the distribution of property accesses locations between the following categories.

- `-O0` refers to the IC in the binary that could not be optimized with DBM because of unsupported IC assembly instruction sequence.
- `-O1` counts the IC in the binary that could only be optimized to the `-O1` level because the second instruction of a pair does not match the expectations.
- `-O2` counts the IC in the binary that have been fully optimized to the `-O2` level.

The numbers atop the bars show the total number of IC that the DBM tried to optimize. A zero means the program does not access any object property and then does not use any inline cache at all. This happens for benchmarks that use arrays exclusively.

Figure 2 shows that more than 96 % of the IC could be modified by the DBM optimization and 17 % could be optimized up to level `-O2`.

☞ This experiment answers *RQ*1 positively: the DBM optimization is able to optimize effectively inline caches.

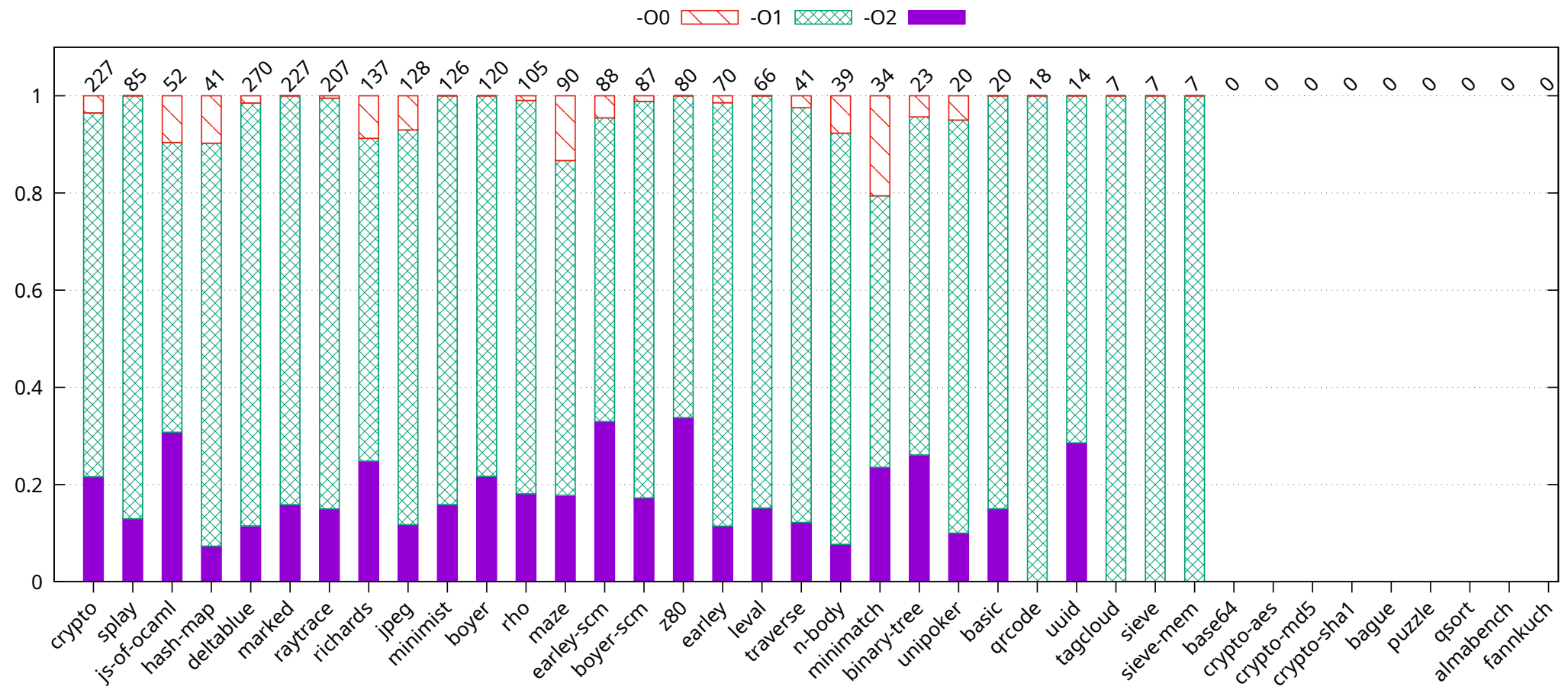

**Figure 2** Effectively applied optimization level for each Inline Cache code point. Numbers atop bars are the executed IC code points count. The bars show the proportion of those accesses that could be fully (`-O2`), partially (`-O1`) or not (`-O0`) optimized.

### 4.4 *RQ*2 What Is the Impact of the DBM on Instructions Count?

Figure 3 shows the number of executed instructions of dynamically optimized code compared to the baseline. Numbers are collected from hardware performance counters with the `perf` Linux kernel tool, counting “`instructions`” events. The optimized version keeps an executed number of instructions very close to the `-O0` version.

Except 2 outliers on Zen 3 (`earley` and `jpeg` which have an increase in instruction count of 1.74 % and 3.33 %), all benchmarks have a variation of less than 1 % (0.01 % on average). Those two outliers can be explained by the insertion of harmless `nop` instructions for padding.

Since programs generated by `hopc` are multithreaded for allocation and garbage collection purposes, not all executions are exactly identical depending on the system's scheduler. From one execution to another, the number of executed instructions are not strictly identical but the differences are small enough to be considered statistical errors.

☞ This experiment answers *RQ*2 by showing that the DBM optimization changes the number of executed instructions marginally.

### 4.5 *RQ*3 What Is the Impact of the DBM on Memory Read?

Figure 4 shows the number of data cache loads (*i.e.*, the memory accesses) per optimizations as measured by the `perf` Linux kernel tool, counting “`L1-dcache-loads`” events. It shows that DBM reduces by 1.5 % on average the number of memory reads, with a peak of more than a 10 % reduction on several benchmarks, for each tested architecture. This was the main aim of the optimization.

☞ This test answers positively *RQ*3: the DBM is an effective technique for reducing the number of the memory reads executed by IC sequences.

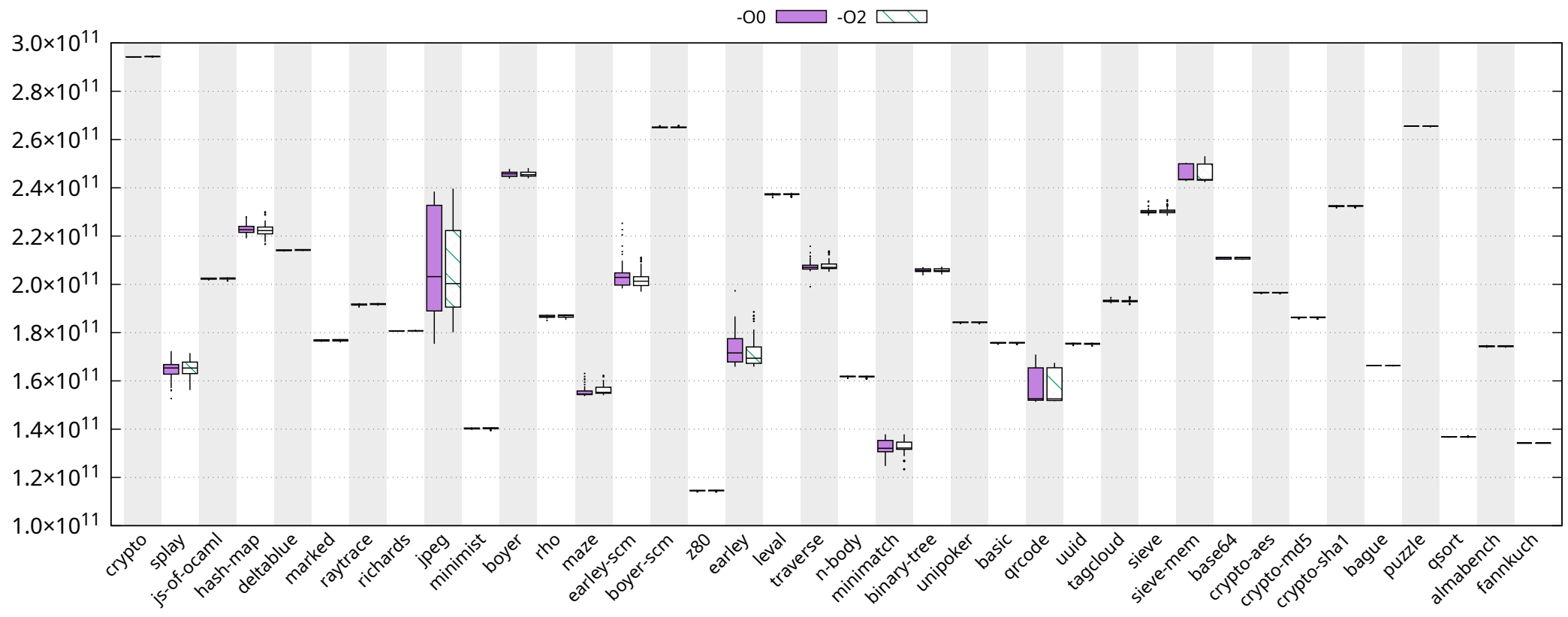

**(a)** On Golden Cove cores.

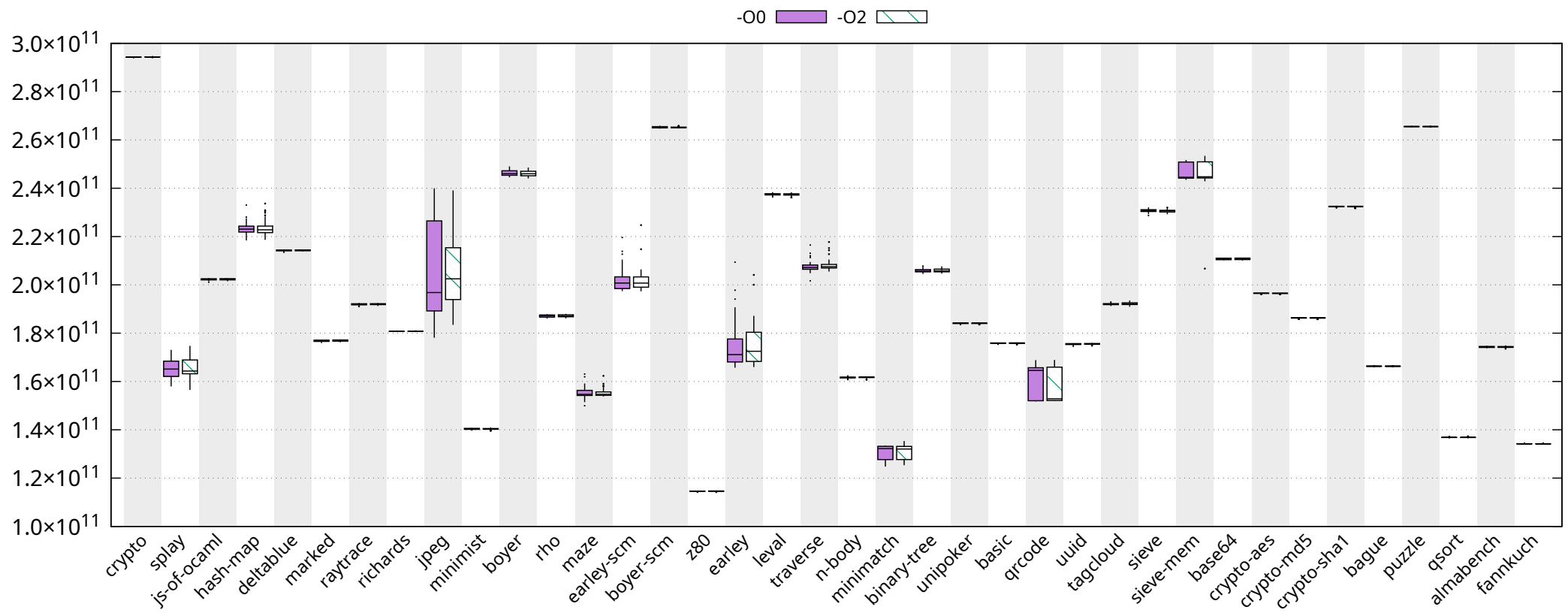

**(b)** on Gracemont cores.

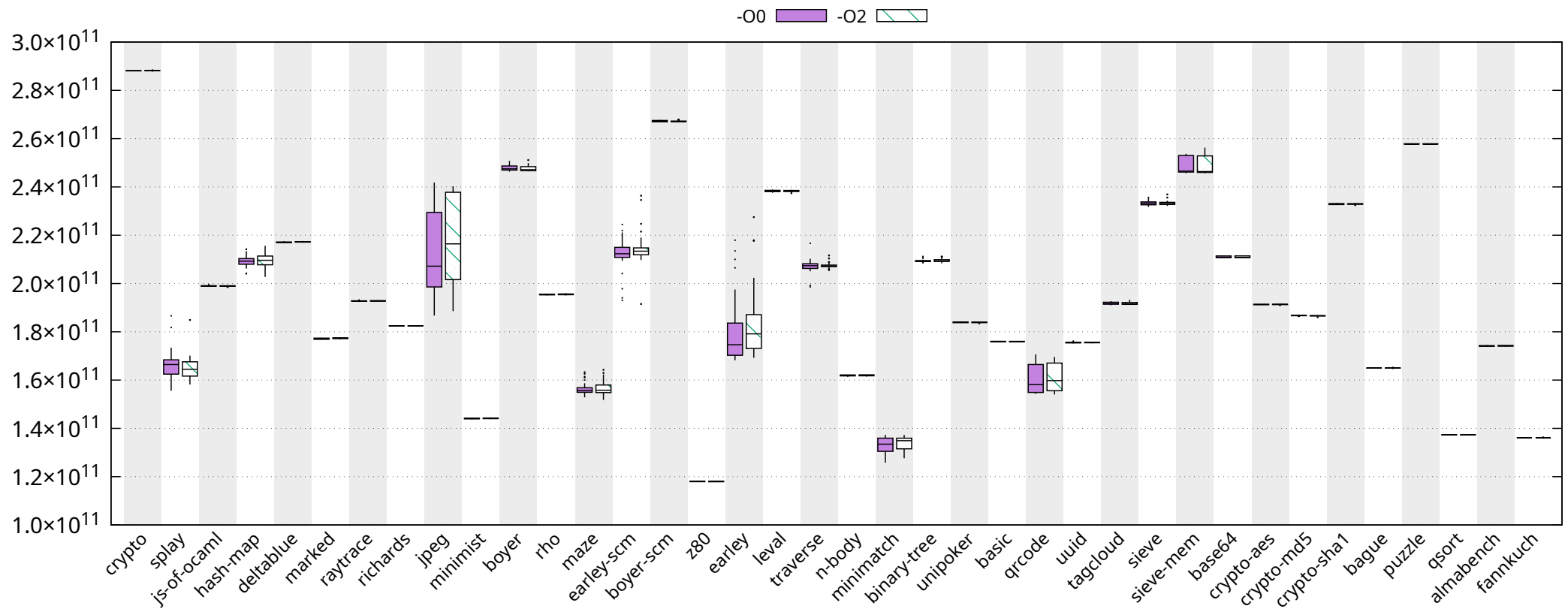

**(c)** on Zen 3 cores.

**Figure 3** Instruction count when applying the optimization compared to `-O0`. Lower is better. Linear scale is used.

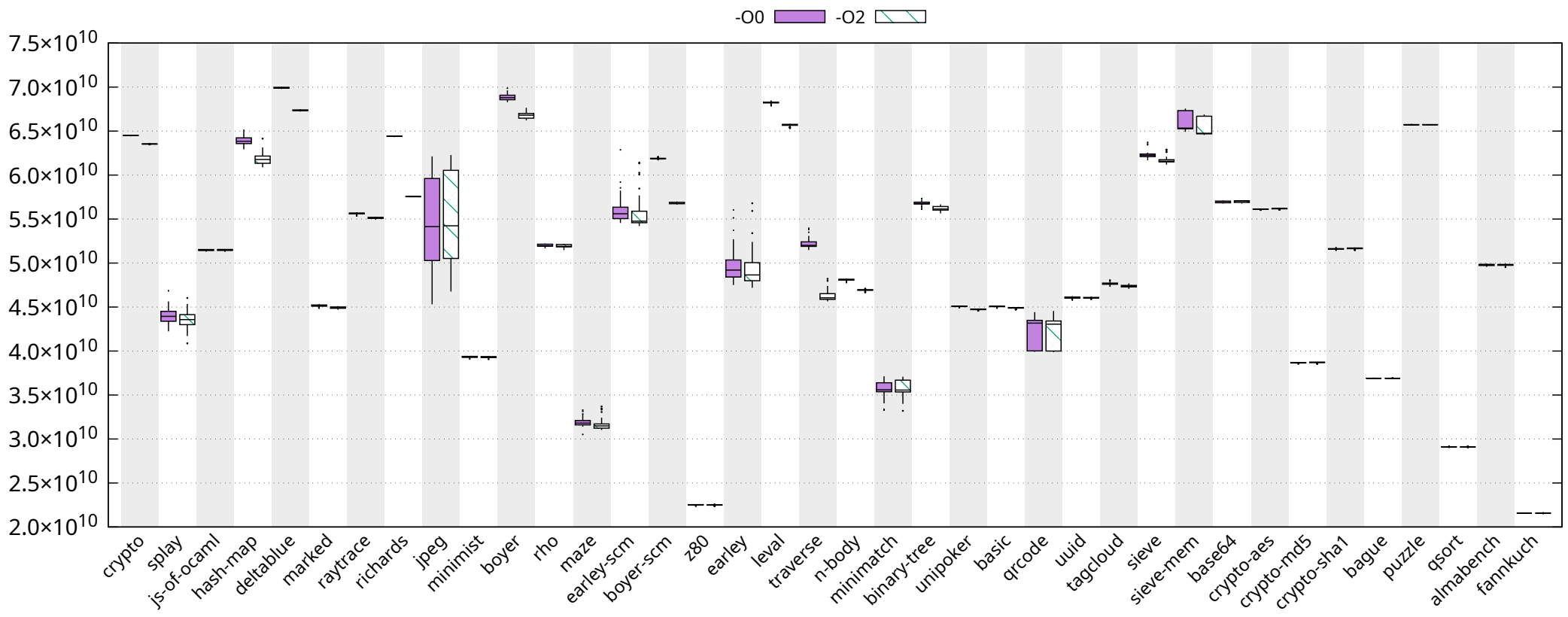

**(a)** On Golden Cove cores.

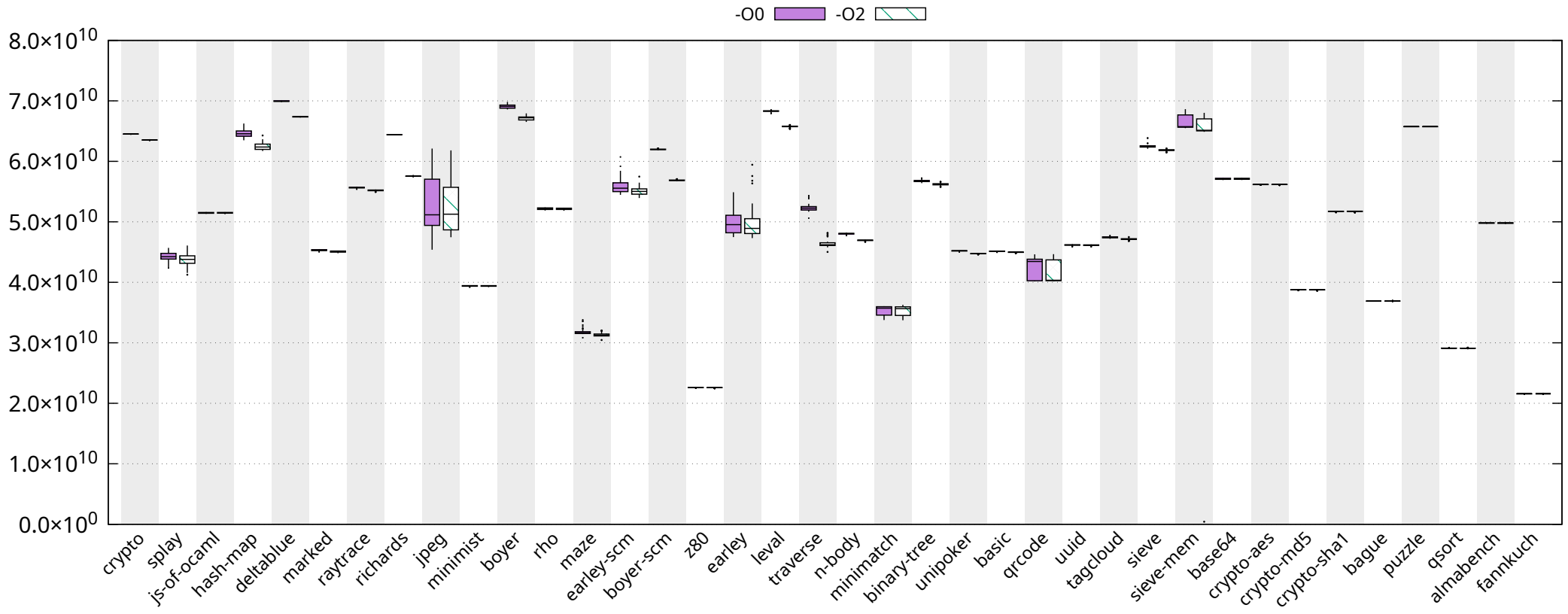

**(b)** on Gracemont cores.

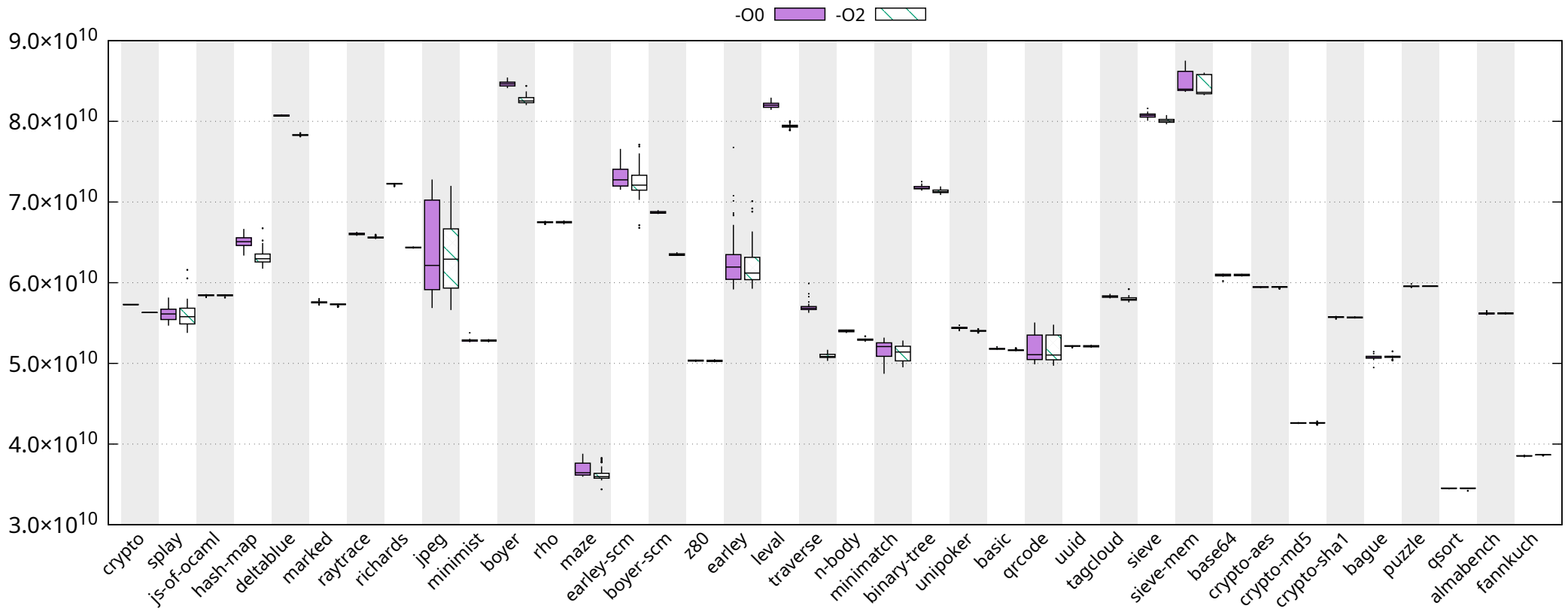

**(c)** on Zen 3 cores.

**Figure 4** Data cache loads when applying the optimization compared to `-O0`. Lower is better. Linear scale is used.

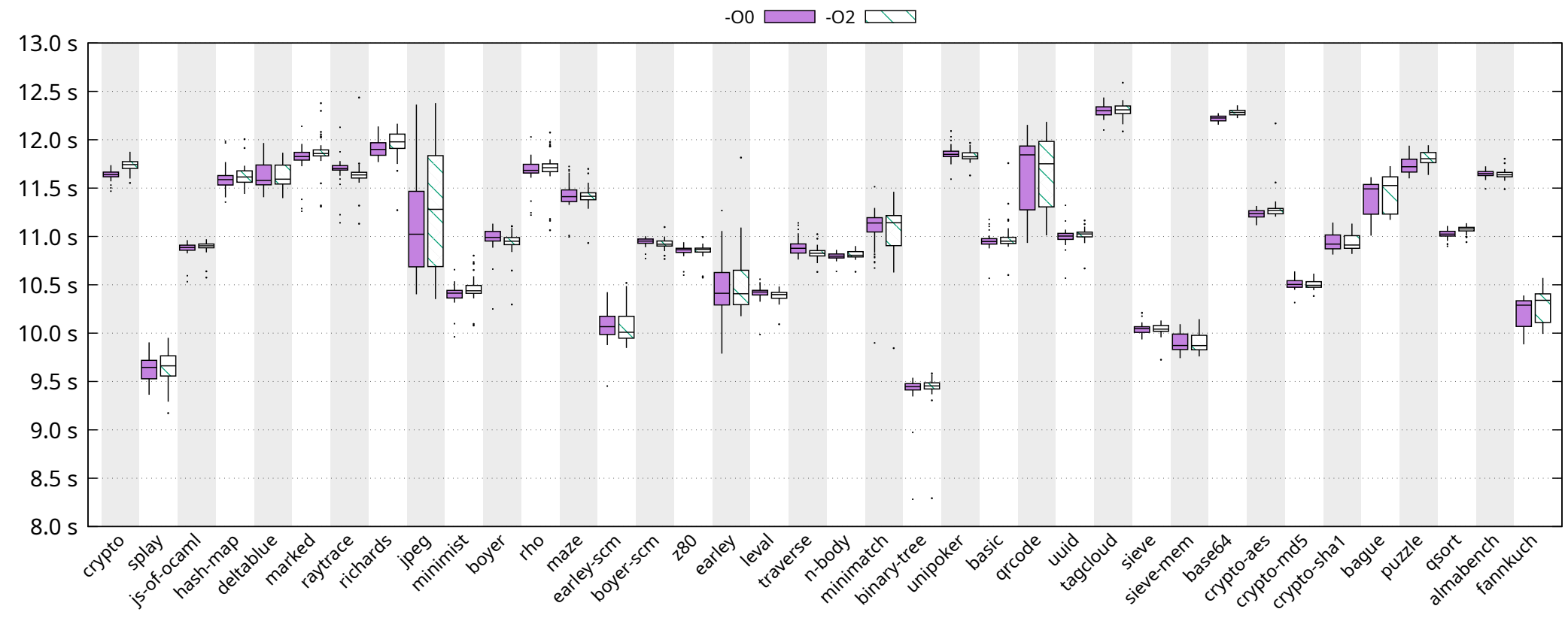

**(a)** On Golden Cove cores.

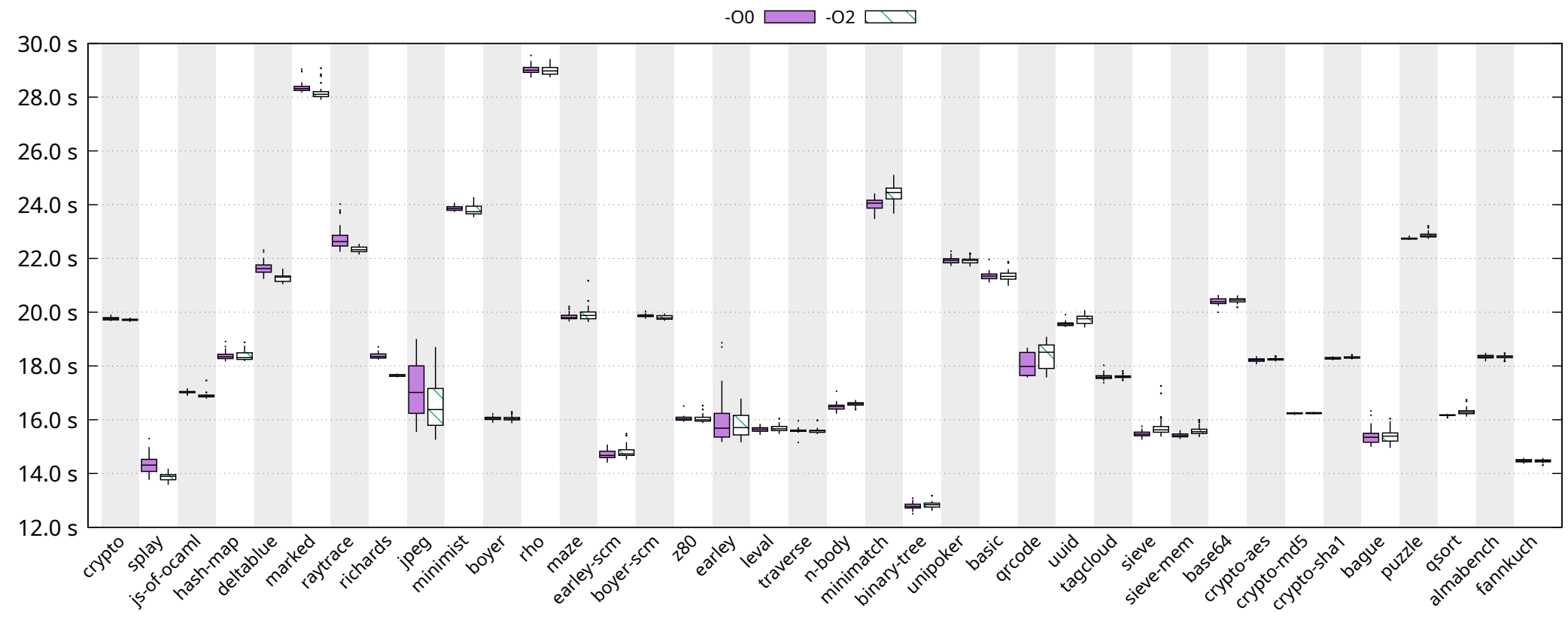

**(b)** on Gracemont cores.

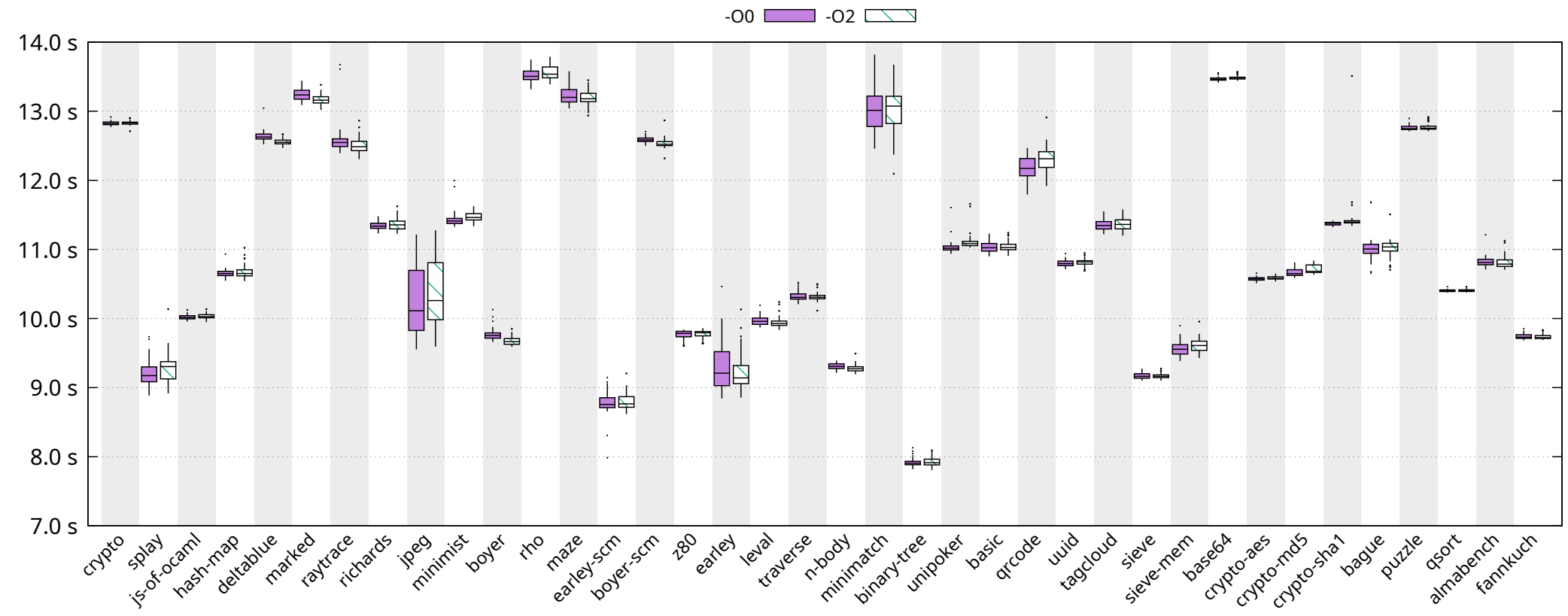

**(c)** on Zen 3 cores.

**Figure 5** Execution time when applying the optimization compared to `-O0`. Lower is better. Linear scale is used.

### 4.6 *RQ*4 What Is The Overall Performance Impact on the DBM Optimizations?

Figure 5 shows the impact of the DBM optimization on the benchmark execution time. At best, it speeds up programs by 4.0 % (with `richards` on Gracemont cores). At worst, it slows them down by 1.8 %. On average, there is a 0.03 % slowdown. According to the literature [18], these minor variations are too small to be considered as a significant execution time difference.

☞ This experiment answers surprisingly *RQ*4: the DBM optimization presented in this paper has no impact on execution speed.

### 4.7 In-depth Study of the Execution Speed

As shown in Sections 4.5 and 4.6 the reduction of memory reads of the DBM does not yield faster executions. In this section, we present the experiments we have conducted to explain this paradox.

Modifying the instructions of the running program is complex. It requires the collaboration of the operating system and the micro-architecture (Section 3.3). Applying DBM to optimized C generated code, it needs a non trivial algorithm for localizing the addresses of the instructions to be modified (Section 3.2). We evaluated the global cost of the operations required to apply the DBM to understand if they counter balance the benefit of executing faster inline caches.

For each of the benchmarks used, we measured *when* the modifications of the executed code take place and *how* frequently they happen, see Figure 6. We observe three distinct execution behaviors:

- DBM only happens during a *warmup* phase that lasts for a couple of milliseconds at the beginning of the execution. Most benchmarks follow this pattern. For instance, the `deltablue` benchmark executes all its modifications in the first milliseconds of its execution and the rest of 12.5 seconds execution uses the optimized code.
- The vast majority of DBM happen during the *warmup* phase but a few residual modifications apply at the end of the execution when the benchmark is terminating, for example to output some pieces of information. For instance, the `maze` benchmark executes 70 modifications during the first milliseconds, then runs for about 11 seconds of optimized code, then executes 5 more modifications at the very end.
- One benchmark keeps modifying the executed code all along its execution.

Let us consider for instance the `deltablue` or `maze` benchmarks. Figure 5 reports no acceleration for these benchmarks when enabling the DBM optimization. Figure 4 shows that they execute extensively optimized IC during their executions and reduce the memory access count. In addition, Figure 6 shows that they apply the modifications at the very beginning of the execution. Finally, Figure 7 shows the execution time for benchmarks tailored for the shortest time possible (limited by a benchmark iteration). It tells us that the overhead imposed by the DBM is less than 10 ms, which is marginal when compared to their whole execution times (more than 10 s, see Figure 5). If there was a significant benefit in running the optimized IC, we should observe a global acceleration for long executions. This is even more obvious for `leval`

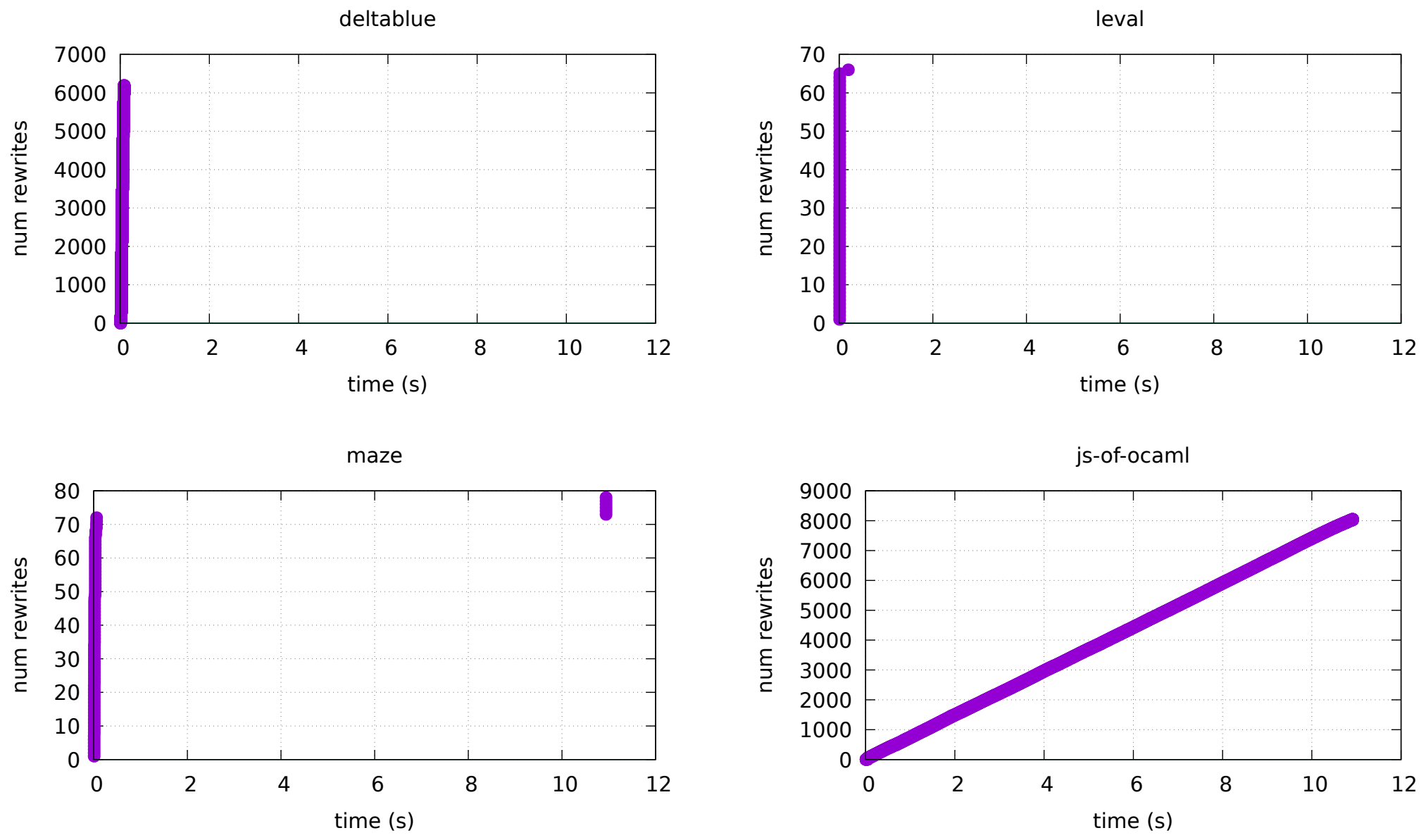

**Figure 6** When and how frequently dynamic rewritings of instructions occur during execution. The horizontal axis is the time in execution (remember that all benchmarks are calibrated to execute in about 10 s). The vertical axis is the number of modifications applied since the beginning of the execution. For instance, this graph tells us that the `js-of-ocaml` executed in between 4000 and 5000 modifications after 6 seconds of execution.

that only modifies the code 65 times during its all execution but the same reasoning apply to all the other benchmarks. As we do not observe any such thing, there can only be one conclusion:

☞ There is no significant benefit in running the optimized IC when compared to running the unmodified IC generated by C compilers.

## 5 Related Works

DBM is an old optimization technique. It was already used in the 80's by efficient implementations of dynamic languages (Smalltalk [10], Self [8]). The expertise developed at that time is still used by implementors of modern JIT compilers (as shown in Section 2) in particular for IC [23] optimization.

DBM has been in use for decades, for various purposes. Wenzl *et al.* [31] and Hazelwood [13] cover a wide variety of uses and techniques. When they write about performance, they focus on reducing DBM overhead, for example in instrumentation context, where performance is already expected to be lowered. Following Wenzl *et al.*'s survey [31], DBM purposes can be classified as follows.

**Emulation.** Emulators are pieces of software meant to reproduce the semantics of a program, usually on a different instruction set, or an older version that does not

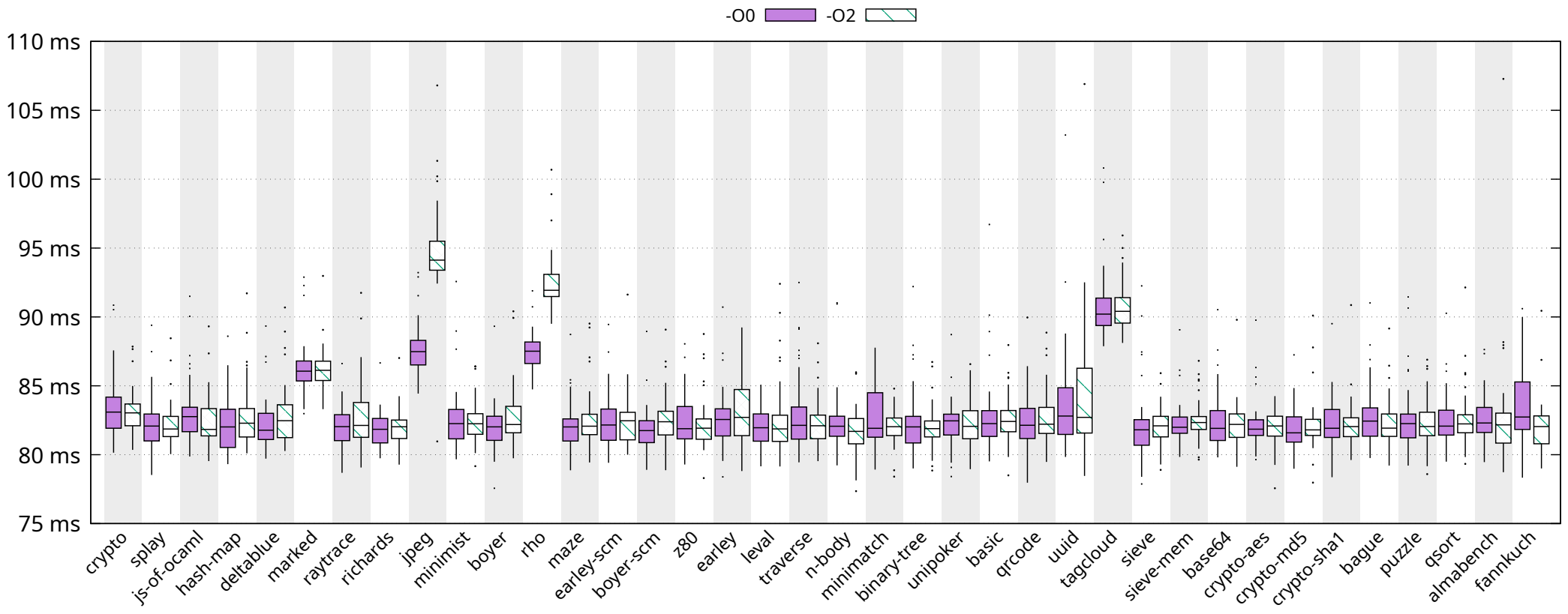

**(a)** On Golden Cove cores.

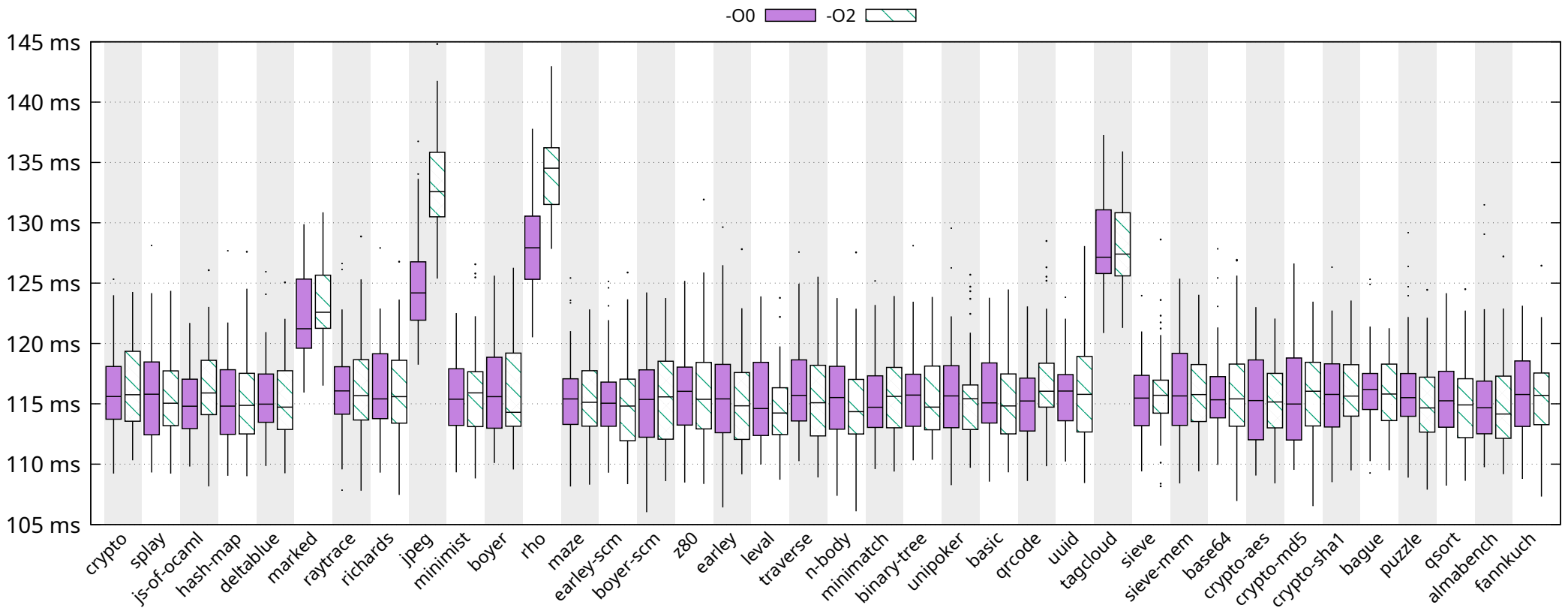

**(b)** on Gracemont cores.

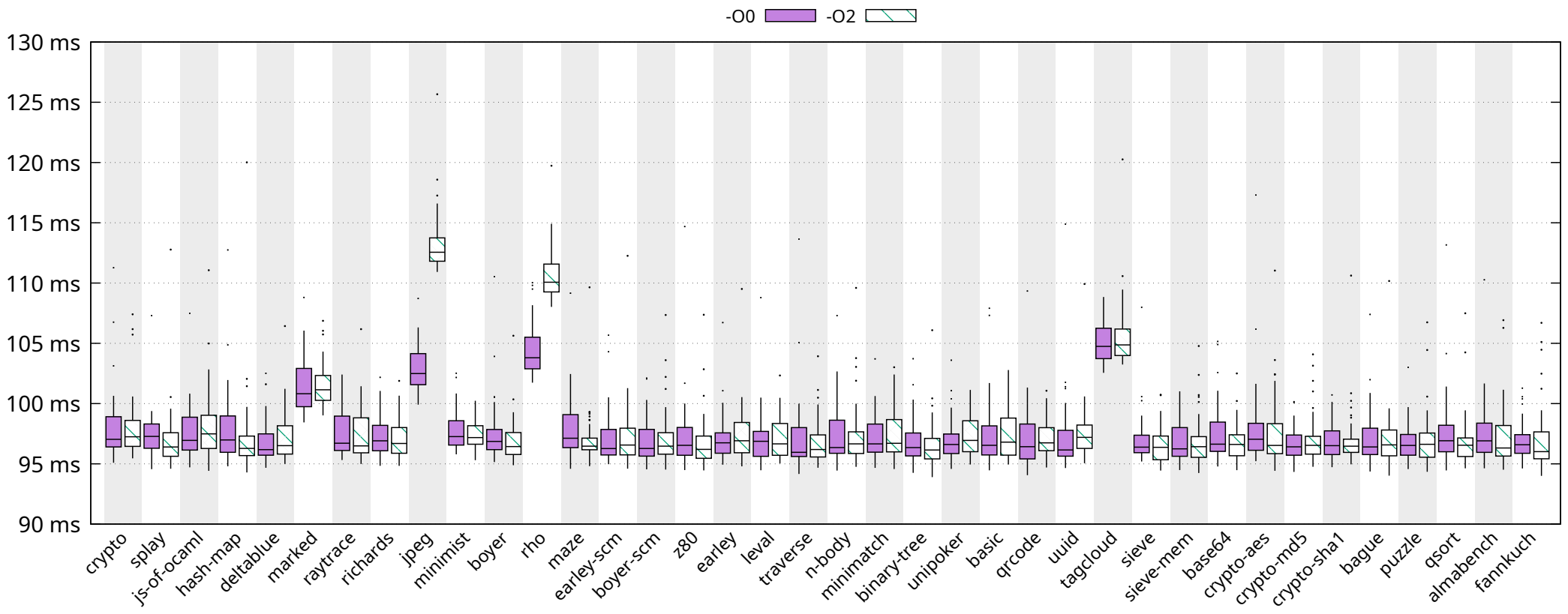

**(c)** on Zen 3 cores.

**Figure 7** Execution time when applying the optimization compared to -O0. Iteration count was tailored to try to reach 50 ms. Lower is better. Linear scale is used.

support all instructions. They can be easily implemented as interpreters, but dynamic binary rewriting delivers much better performance. QEMU is representative of such a software [5].

**Observation.** Programs can be observed using DBM. Pin [17] and Valgrind [19] work as JIT compilers to add instrumentation code around the original program instructions. GDB [29] dynamically adds `int3` instructions to interrupt program to balance between execution speed and code analysis.

**Optimization.** Bala *et al.* presented Dynamo [4], a user-land dynamic binary optimization system. When an address in branched to a high number of times, its code is traced, then optimized. Their main optimization consists in re-ordering instructions and removing branches. They also apply conventional optimizations such as constant propagation, loop unrolling and redundant loads removal. Hallou *et al.* [12] noticed that binary programs typically do not exploit instruction sets to their fullest extent because they target older processors (mainly for portability purposes). They illustrate this with the SIMD vector instruction families and they show how DBM can translate x86 SSE idioms to its successor AVX, thereby doubling the potential throughput of loops. Ana-Pparakkal *et al.* explored different ways DBM can improve performance: first they proposed to rely on runtime information to apply function specialization [3] and then investigated how program optimization can be applied at runtime [2] be repeatedly invoking the compiler on hot pieces of code. Somewhat related to our approach, Nuzman *et al.* [20] proposed to recompile C/C++ programs at runtime. The main difference is in the fact that they embed the compiler intermediate representation in so-called *fat binaries*, while we optimize highly focused sequences of instructions.

**Hardening.** DBM is a convenient mechanism to harden programs, in particular unprotected ones. As an example, control-flow integrity ensures that no attack is made on control flow. Payer *et al.* [21] translates dynamically binaries to add checks of indirect branches destinations against a destination *allow-list*. Replacing indirect branches by direct branches is the approach taken by Le Bon *et al.* [16].

DBM comes in two main shapes: in-place modifications and writing entire code parts. The latter is used way more often than the former because DBM applications often requires inserting instructions that do not fit in the available space or the target binary has some incompatibilities with the source one, for example targeting a different ISA [5].

In-place DBM has been used for performance purpose. It is often used to remove instructions and branches that are not executed [6]. As seen in Section 2, JIT compilers also use in-place DBM, but have no static baseline to compare to.

Reliably measuring program performance is difficult. General purpose modern processors achieve high performance thanks to extremely complex microarchitectures and highly dynamic behaviors (dynamic voltage and frequency scaling – DVFS, Turbo Boost, simultaneous multithreading…). As a consequence, even seemingly innocuous aspects can lead to wrong conclusions about system performance, as shown by Mytkowicz *et al.* [18].

Curtsinger and Berger presented Stabilizer [9], a research framework to get reliable performance measurements. It works by randomizing heap and stack allocations and by dynamically stopping the program to relocate the executed code every half second. This method is incompatible with the DBM technique presented in this paper.

Pichler *et al.* [22] proposed a way to hybrid between Ahead-of-Time and JIT compilers. Functions can be either compiled statically or use a JIT compiler. For now, the functions to be statically compiled have to be selected manually.

## 6 Conclusion

In this paper we present a technique we developed to enable dynamic optimization of static code produced by C compilers. We implemented and tested the optimization in the context of `hopc`, a static JavaScript-to-C compiler, for which it improves the performance of *inline caches* (IC), a common technique used to optimize object accesses. By using *dynamic binary modification* (DBM) of the assembly sequences C compilers generate, `hopc` eliminates up to two memory reads by IC, yielding executable codes that are as efficient as those generated by JIT compilers.

We measured the impact the optimization and observed that while it eliminates memory reads, as intended, it does not accelerate executions. We conducted several experiments to explain this behavior. We isolated the intrinsic cost of modifying the binary code of programs and showed that itself it cannot explain the lack of acceleration. We eventually discovered, that the optimized IC sequences are actually no faster than un-optimized ones. The experiments we present in this paper show that in the context of IC sequences, the micro-architectures already executed the extra memory reads of the C sequences early, and there is no benefit in removing them.

None of the tools we tried, including performance analysis tools that use hardware performance counters and micro-architectural code analyzers [1], was able to predict or even unveil this hardware optimization. Its only by running many experiments that we have been able to isolate and observe the hardware optimization that removes the extra memory we were chasing with our software optimization.

## A Statistical Significance of Section 4.2

In Section 4.2 we concluded: "This experiment shows that adding *unexecuted* code (`-O0` versions) to programs impacts their performance by ±5 %." The current appendix aims at ensuring statistically the validity of this result by showing that it is consistent between executions and not a result depending on measurement noise.

The sample size being 100 for each measurement, the distribution can be supposed to be normal. No hypothesis can be made about a variance equality between compared samples. The measurements still satisfy the conditions to run Welch's t-test. To be comparable to any possible other work, we pick $\alpha = 0.05$, which is a usual value. If Welch's t-test p-value is under this threshold, we can conclude any measured performance difference is significant.

Let us take the benchmarks with the highest speedup and the highest slowdown.

- `earley` got a 5.28 % speedup on Golden Cove cores.
- `crypto-md5` got a 5.17 % slowdown on Gracemont cores.

Testing `earley` on Golden Cove with Welch's test gave a $1.15 \times 10^{-15}$ p-value. Testing `crypto-md5` on Gracemont with Welch's test gave a $5.76 \times 10^{-114}$ p-value. Both those values are below 0.05, which allows us to conclude that those two results are statistically significant: there are speedups and slowdowns of more than 5 %.

## B Cache Misses

Figure 8 shows the L1 data cache misses of benchmarks when applying the optimization or not. As Gracemont cores do not provide hardware counter for such metric, it is not present on this figure.

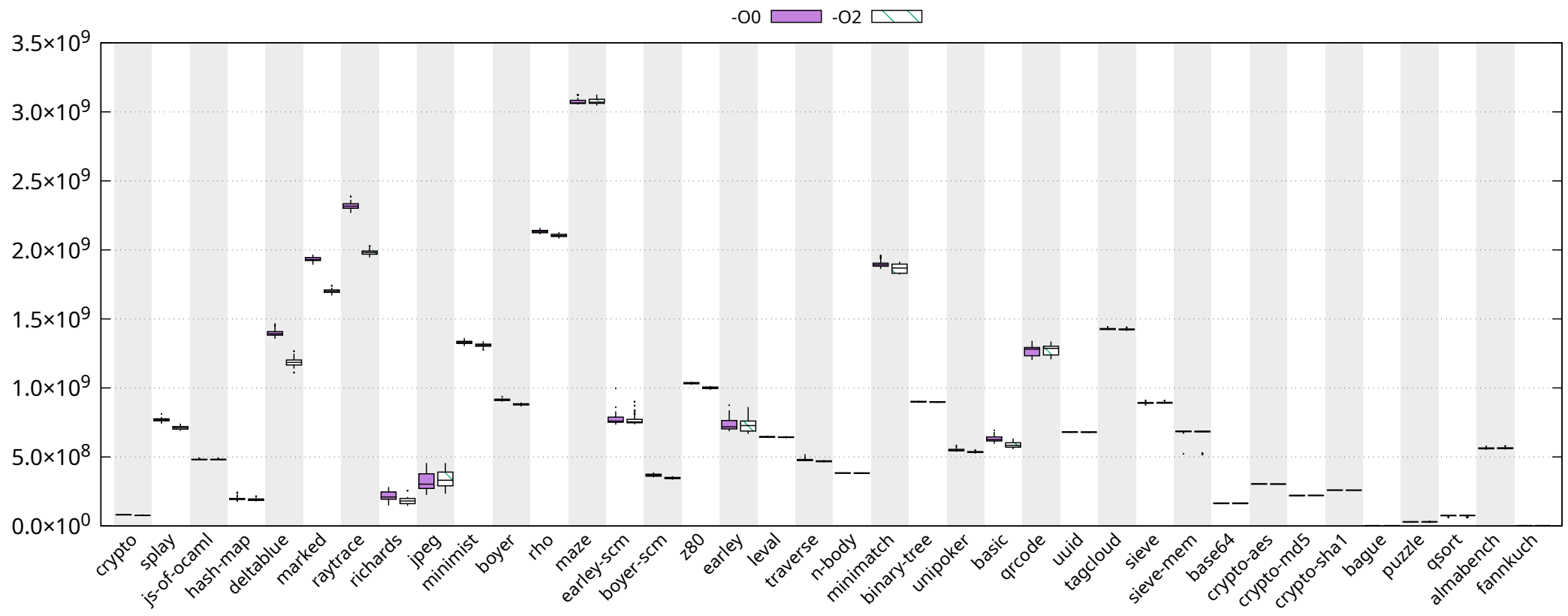

**(a)** On Golden Cove cores.

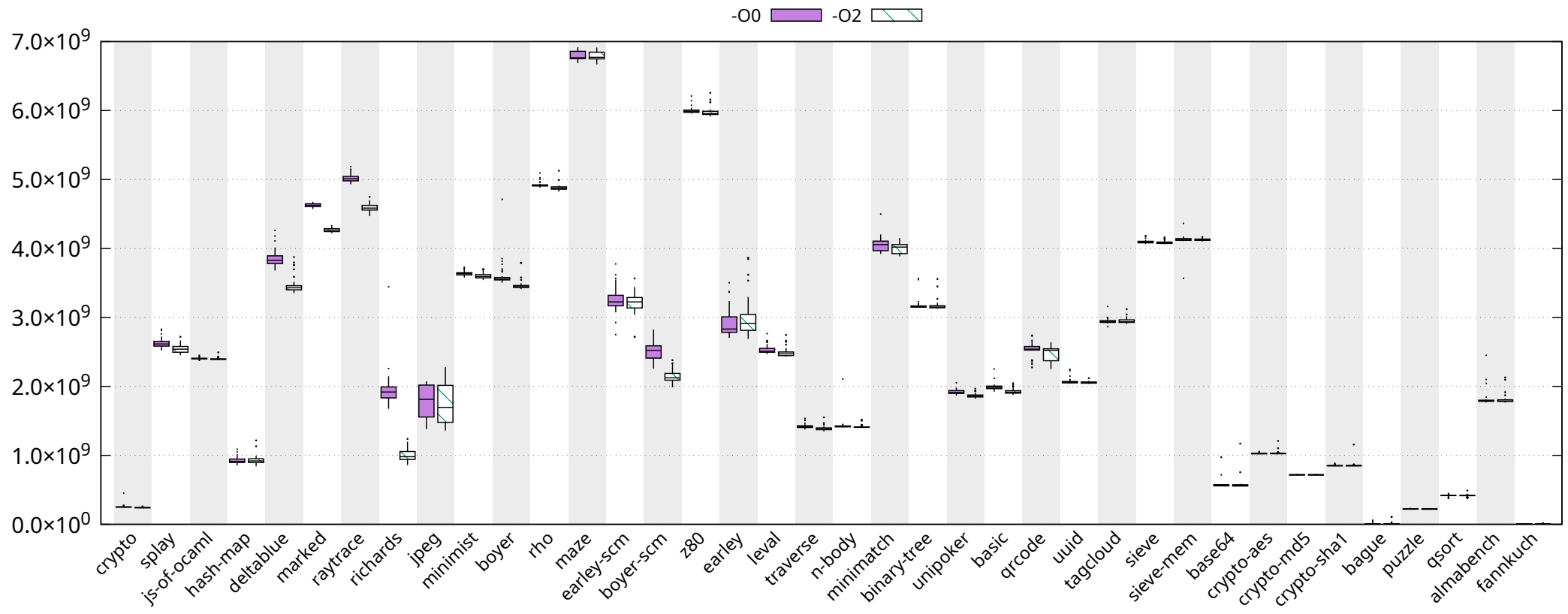

**(b)** on Zen 3 cores.

**Figure 8** L1 data cache misses when applying the optimization compared to `-O0`. Identical conditions to Section 4.6 were applied. Lower is better. Linear scale is used.

## About the authors

**Aurore Poirier** is a PhD student at Inria. Contact them at aurore.
poirier@inria.fr.
https://orcid.org/0009-0001-5155-2580

**Erven Rohou** is a Senior Researcher at Inria and the head of the PACAP Inria project-team. Contact him at erven.rohou@inria.fr.
https://orcid.org/0000-0002-8060-8360

**Manuel Serrano** manuel.serrano@inria.fr
https://orcid.org/0000-0002-5240-1610